\documentclass[10pt]{ismd08}
\usepackage{graphicx}
\usepackage{cite,./mcite}

\begin{document}
\title{Bose-Einstein or HBT correlations in high energy reactions}
\author{T. Cs\"org\H{o}
\protect\footnote{\ \ e-mail: csorgo@rmki.kfki.hu}
 }
\institute{MTA KFKI RMKI, H-1525 Budapest 114, P.O.Box 49, Hungary 
}
\maketitle
\begin{abstract} 
Concepts of  thermalization and hydrodynamical behavior are applied from time to time
to  e$^+$+ e$^-$ , hadron+hadron and heavy ion collisions. These applications are scrutinized 
paying attention  to particle multiplicities, spectra, and 
Bose-Einstein correlations in particular.
Can hydrodynamics describe these data?
\end{abstract}

\section{Introduction}
In 2008, the speakers of the International Symposium on Multiparticle Dynamics 
were given a quiz of 18 questions, that were compiled by Hannes Jung and G\"osta Gustafson,  
the Chair and the Co-Chair of this meeting~\cite{quest}.
My goal is to discuss three of these problems:
\begin{enumerate}
\item{} Can thermal \& hydrodynamical models describe 
$e^{+}$+$e^{-}$, $h$+$p$ and $A$+$B$ reactions?
\footnote{ Instead of the originally given $e^-$+p problem, 
let me discuss soft hadron-proton collisions,
for clarity. }
\item{} What heavy ion physics can learn from 
$e^{+}$+$e^{-}$, $h^-$+$p$ and $p$+$p$ collisions?
\item{} How can correlations be used to determine the size of the interaction region and the characteristics of phase transitions?
\end{enumerate}

These questions are related to 
Bose-Einstein correlations, that appear due to the symmetrization of hadronic final states
for the interchange of identical bosons, and are also known by other names,
for example  Hanbury Brown -- Twiss or HBT correlations in heavy ion collisions, 
intensity interferometry, or intensity correlations~\cite{Csorgo:1999sj}. 
These correlations are also tools of femtoscopy,
because they are used to measure length scales on the femtometer scale~\cite{Lednicky:2002fq,Lisa:2005dd,Bekele:2007ee}.

\section{The shortest film ever made: $e^+$ $e^-$ collisions at LEP}
\label{sec:movie}
In $e^+$ $e^-$ collisions, Bose-Einstein correlations 
were used to record the fastest film
ever made: the formation of a ring-like, {\it non-thermal} source in the 
transverse plane of jet production, a process that ends 
in less than $10^{-23}$ seconds~\cite{Novak:2006sw,Novak:2008PhD}.
Can {\it thermal} models describe multiplicities, spectra and correlations in these
collisions? 

A number of recent papers consider the possibility 
of thermal particle production in $e^+$ $e^-$ 
reactions. Two recent, interesting examples are
refs.~\cite{Andronic:2008ev} and ~\cite{Becattini:2008tx} , 
that present thermal model fits to these data with similar level of statistical 
significance but with very different physics conclusions.
A model cannot be excluded with the help of mathematical statistics
if its confidence level is CL$ \ge  0.1 \%$, thus the probability that the model describes the data is at least one in thousand.

Fig. 1 of ref.~\cite{Andronic:2008ev} is a very beautiful plot indicating intriguing
similarities between particle abundances in 
$e^+$ $e^-$ at $\sqrt{s_{NN}} $ = 91 GeV 
and thermal model calculations. The fit quality is
characterized by a  $\chi^2$/NDF   = 631/30. The 
corresponding confidence level is CL = 1.1 10$^{-111}$ \%.
This confidence level is  an extremely small positive number and so 
the probability that the thermal particle production describes this data set is 
practically zero~\cite{Andronic:2008ev} . 
These authors also observe 
and point out correctly that the statistical or thermodynamical description of these
data fails completely at the high level of present experimental precision.
Approximate qualitative agreement between thermal particle production and data
can only be obtained if the relative errors on these data
are magnified to about 10 \%  ~\cite{Andronic:2008ev}.
Their conclusion
can be contrasted to other manuscripts, that claim 
that  a thermodynamical or statistical description of particle multiplicities
in $e^+$ $e^-$ reactions is possible. For example, the same data 
set was analyzed in ref.~\cite{Becattini:2008tx} ,
using a slightly different  thermal model description
 and the quality of their fit is given in their Table V  as $\chi^2/$NDF = 215./27 . 
The corresponding confidence level is CL = 3.4 10$^{-29}$ \%
hence the probability that this thermal model describes particle
abundances in electron-positron annihilation is practically zero.  
When the analysis is restricted to
include only those 15 resonances in the fitting,
whose width is less than 10 MeV, the same thermal model description yields 
$\chi^2/NDF$ = 39./12 . The confidence level of this fit is still CL = 1.1 10$^{-2}$ \% ,
which is many orders of magnitude improvement,
but still an order of magnitude 
less than the conventional threshold of acceptance,
CL = 0.1 \%.  Thus thermal particle production models do not describe the multiplicities
of elementary particles of $e^+$ $e^-$ at $\sqrt{s_{NN}} $ = 91 GeV 
in an acceptable manner. 

Two important and well known features of hadronic spectra  also
disagree with a thermal, statistical picture of particle production. 
The observation of jets (2 and 3 jet events at this energies) can be contrasted to 
the lack of preferred direction in the initial conditions and in a thermal
picture of particle production. Perhaps the thermal picture can be limited to the 
transverse direction?  The power-law tail the transverse momentum spectra, 
which can be explained in terms of perturbative QCD processes and jets decaying to jets to jets
and in particular the correlations among these jets are inconsistent with 
a thermal and/or a hydrodynamical interpretation, that lead typically to exponential spectra. 
Furthermore, generalized thermal models that describe the spectra 
cannot naturally interpret the correlation structures observed 
in two and three jet events which are basically energy momentum conservation laws  
and have a trivial interpretation in partonic picture, the emission of quark and gluon
jets in  perturbative QCD.

\begin{figure}[t]
  \begin{center}
    \includegraphics[width=15.5cm]{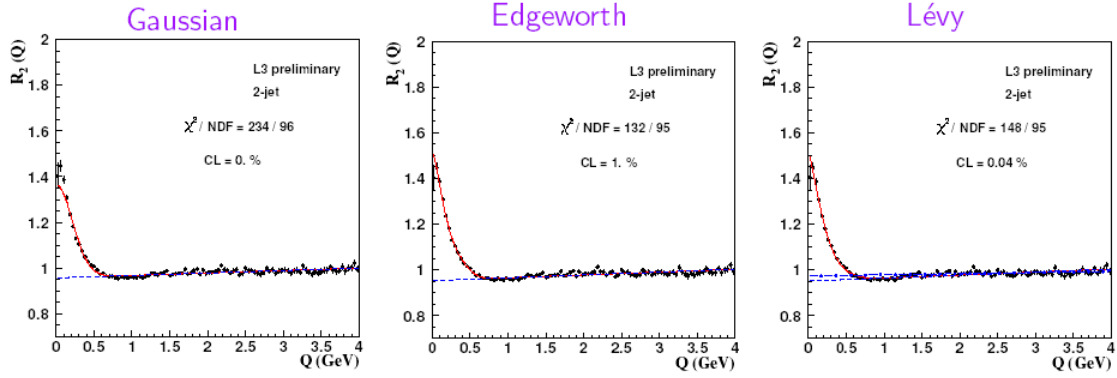}
  \end{center}
\vspace*{-0.8cm}
  \caption{
Comparison of Gaussian (left), Edgeworth (middle)  and L\'evy fits to L3 Bose-Einstein 
correlation functions. The  1+ positive definite forms, Gaussian and L\'evy do not have a 
statistically acceptable level, CL $< 0.1   $\%. 
The Edgeworth expansion has an acceptable CL = 1 \% and 
describes the dip using a 1+ non-positive definite expression.
}
 \label{f:Gauss-Edge-Levy}
\end{figure}

Bose-Einstein correlations are more subtle features of two-particle distributions.
They carry information on the space-time structure and on the chaotic or coherent nature of
particle emitting sources.  Recently measured Bose-Einstein correlations 
disagree qualitatively with the hypothesis that the produced
particles are emitted from a thermal or hydrodynamical source in $e^+$ + $e^-$
reactions, because in thermal models the two-particle Bose-Einstein correlation 
function is always given by a  1 +  positive definite function,
and this constraint is violated by a recent analysis of L3 data~\cite{Novak:2008PhD,Wes:ISMD2008}.With other words, there is no region of two-particle relative momentum space,
where a chaotic (or thermal) picture of particle
production would lead to anti-correlations. 
However, recently analyses L3 data as detailed in ref.~\cite{Novak:2008PhD,Wes:ISMD2008} 
indicate very clearly the existence of a region of anti-correlation: 
if the correlation functions are measured
as a function of  the Lorentz invariant relative momentum variable $Q = \sqrt{-(k_1 - k_2)^2}$,
where $k_i$ stands for the four-momentum of particle $i$, 
a dip is found experimentally in the region of 
0.6 GeV $< Q < $ 1.5 GeV,
as indicated in Fig.~\ref{f:Gauss-Edge-Levy}.  
In this  kinematic range the errors are small. 
This feature is shown in greater details in Fig~\ref{f:Gauss-Edge-Levy}. L3 data from ref.~\cite{Novak:2008PhD}
are compared to a Gaussian fit, $C(Q) = 1 + \lambda \exp( - Q^2 R^2)$, $\chi^2/NDF = 234./96$
and the corresponding confidence level is practically zero. 
A generalization of the Gaussians is given
by the symmetric L\'evy form $C(Q) = 1 + \lambda \exp( - Q^\alpha R^\alpha)$, 
$\chi^2/NDF = 148./95$, CL = 0.04 \%.
This is only a factor of 2.5 below the conventional domain of acceptable
results, but the chance that this form represents the data is only 4 in 10000.
This form however is 1 + a positive definite function. 
The right panel also indicates a linear fit to the long range, 
$Q > 1.5 $ GeV  correlations, shown with a dot-dashed line.  
It describes the data in the fitted $Q > 1.5$ GeV region, and it clearly cuts into 
the "dip" region of the data, located at $0.6 < Q  < 1.5 $ GeV . 
When a L\'evy fit form is enforced, these long range correlations get distorted,
pushed below the dip region by the fit, as indicated by the dashed line in the right panel,
and the overall fit quality is decreased below the limit of acceptability, CL $ < 0.1$ \%.
The best fit is achieved using an Edgeworth expansion, 
$C(Q) = 1 + \lambda \exp( - Q^2 R^2) [ 1 + \kappa_3 H_3(QR) ]$,  
where $H_3(x) $ is the third order Hermite polynomial,
see ref. ~\cite{Novak:2008PhD} for details. This Edgeworth fit
has a statistically acceptable CL = 1 \% and 
describes the dip using  1 + a non-positive definite expression, in a model 
and interpretation independent manner.

The $\tau$-model of ref.~\cite{Csorgo:1990up}
also predicted 
the existence of such anti-correlated regions, 
based on the assumption that 
$e^+ + e^-$ annihilations indeed correspond to point-like collisions hence the produced particles with a given momentum $k^{\mu} $ appear in a direction parallel to their momentum, 
$x^\mu \propto k^\mu$, however with a broad proper-time distribution $H(\tau)$.
This model 
leads to simple fitting forms, 
that improve the description of the data as compared to the model-independent 
Edgeworth expansion method,
CL is increased from 1 \% to 40 \%  ,
and when the fit parameters are required to satisfy the model constraints,
CL is slightly increased to 42 \% , 
see ref.~\cite{Novak:2008PhD} for details.
The parameters of the proper-time distribution
are determined from detailed fits to the L3 Bose-Einstein correlation functions.
This way the proper-time evolution of particle production  is reconstructed
in these reactions, and the following  
points can be made: in 2-jet events, particle emission starts 
just after the collision, so that the most probable value for
$\tau$ is 0.3 fm/c, but this one-sided proper-time distribution has a power-law tail,
corresponding to a one-sided L\'evy distribution with an index 
of stability of $\alpha = 0.42 \pm 0.01$. Using a recently developed
method based on the $\tau$-model~\cite{Csorgo:2008ah},
 even a movie of the space-time evolution of particle emission
can be reconstructed. This movie 
-- the shortest film ever recorded --
practically ends in about 0.3 fm/c.
\section{The smallest ring of fire: $h$+ $p$ and $p$+$p$ collisions }
\label{sec:ring}

In hadron-proton collisions, Bose-Einstein correlations have been used to make a 
snapshot picture of the smallest ring of fire, ever detected: 
the diameter is less than  1 fm or $10^{-15}$ m, but the source seems to be thermal. 
 The ring formation here is a hydrodynamical effect, 
the temperature drops from $T \approx 140 $ MeV in the center 
to nearly zero within about 1 fm radial distance, 
hence a strong pressure gradient builds up. 
However, the experimentally seen transverse flow 
is too week to move the matter away from the surface,
hence a pile-up at the surface, a fire-ring is found~\cite{Csorgo:1999sj,Agabab:1998ehs}.
A similarly hydrodynamical ring of fire formation due to large temperature gradients 
and small transverse flows can be inferred from a simultaneous analysis of 
single particle spectra of pions, kaons, protons and STAR preliminary 
Bose-Einstein or HBT correlation radii of pion pairs in $\sqrt{s_{NN}} = 200 $ 
GeV p+p collisions at RHIC~\cite{Csorgo:2004id}.
\begin{figure}[t]
  \begin{center}
\begin{tabular}{cc}
    \includegraphics[width=7.0cm]{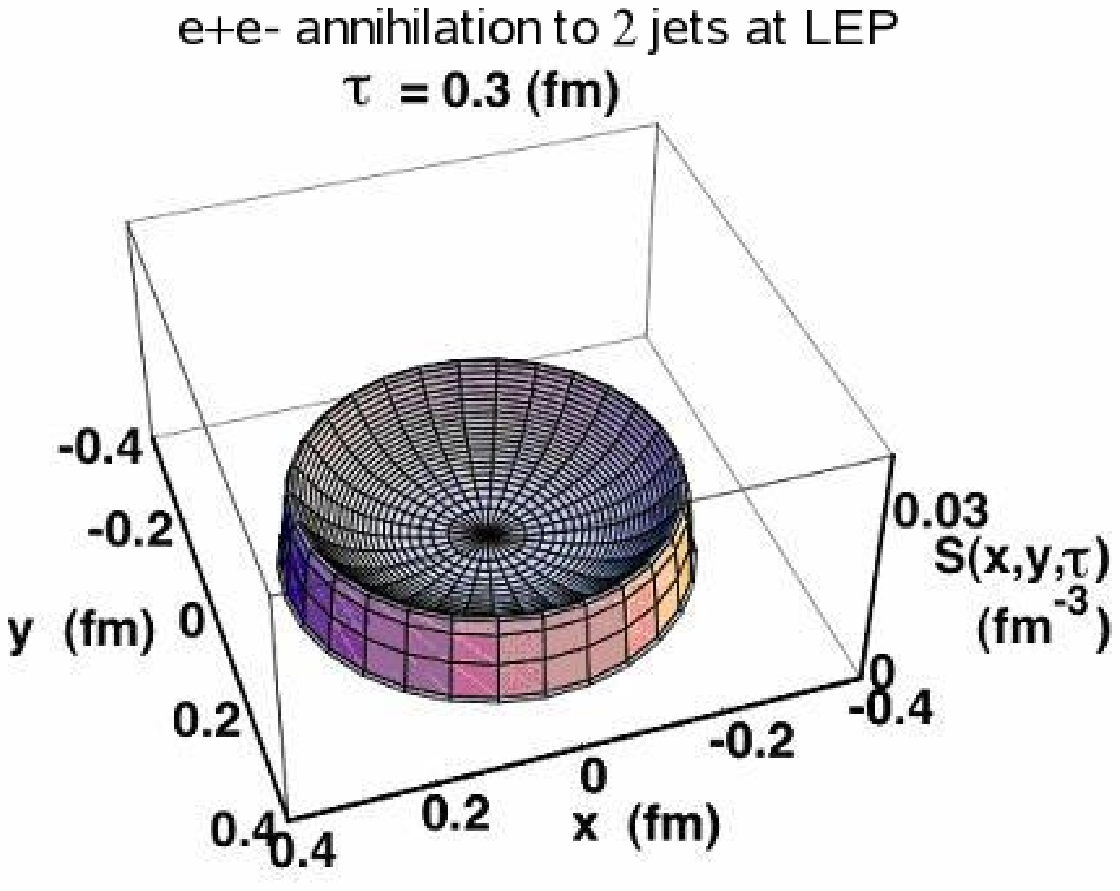} 
&
    \includegraphics[width=7.0cm]{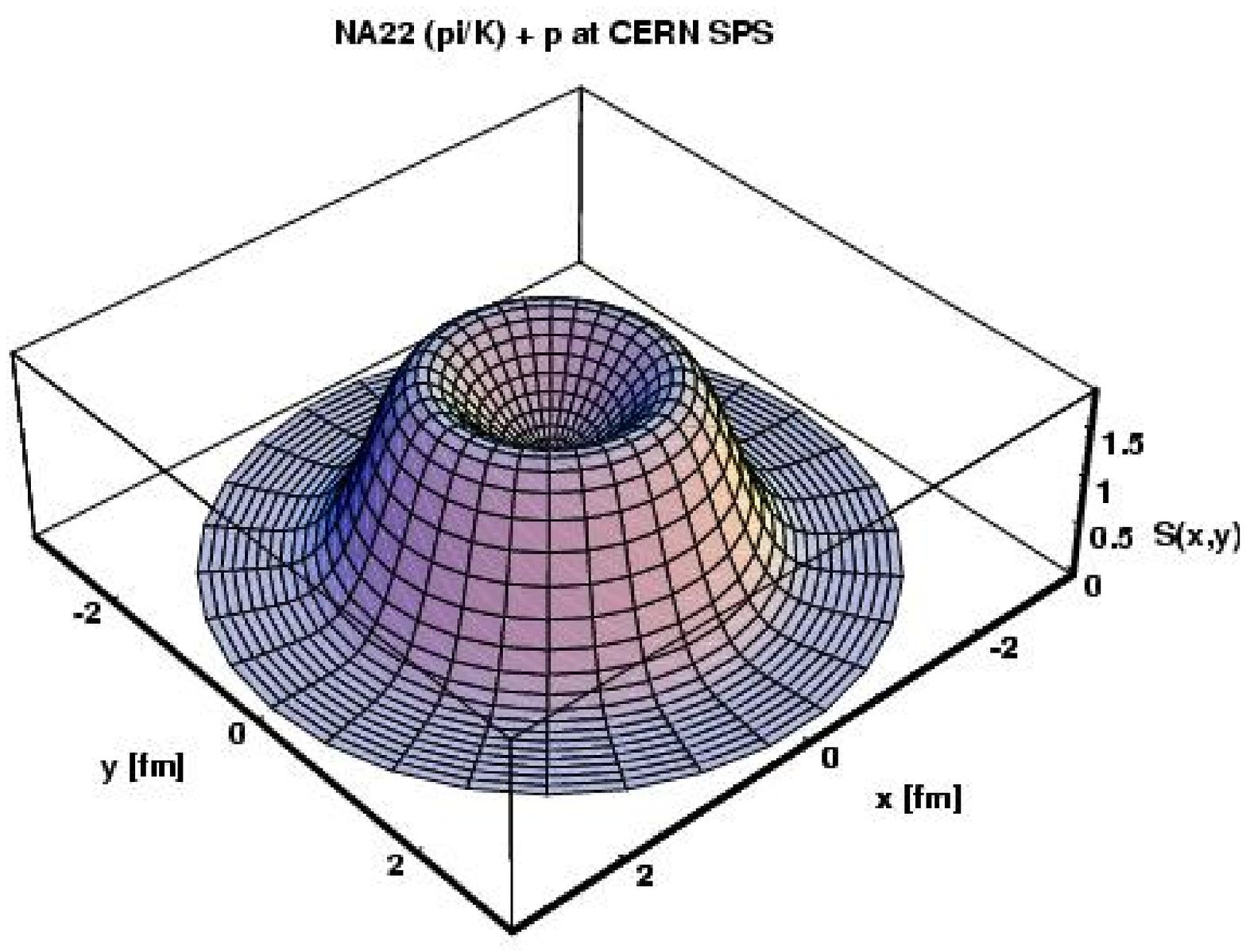} 
\end{tabular}
  \end{center}
\vspace*{-0.5cm}
  \caption{
Left panel: 
A snapshot picture from the reconstructed video of jet formation in the transverse plane of
2-jet events in $e^+$ + $e^-$ annihilation at LEP at $\tau = 0.3 $ fm/c. A non-thermal, expanding ring is observed, the amplitude of the ring diminishes very quickly, while its radius grows 
nearly with the speed of light.
Right panel: 
The reconstructed transverse part of the particle emission function in h+p reactions at CERN SPS as inferred from Bose-Einstein correlations and single particle spectra as 
measured by the NA22 collaboration, describing a tiny ring of fire.
}
  \label{f:e+e-peeblepond}
\end{figure}

\section{The hottest and most perfect fluid: $Au+Au$ collisions at RHIC}
\label{sec:sfluid}
In Au+Au collisions at $\sqrt{s_{NN}} = 200 $ 
GeV at the RHIC accelerator at BNL, 
Bose-Einstein correlation measurements  also yield  snapshot pictures of the hottest and 
most perfect fluid, ever made in a laboratory experiment.

\begin{figure}[tb]
  \begin{center}
    \includegraphics[width=10.0cm]{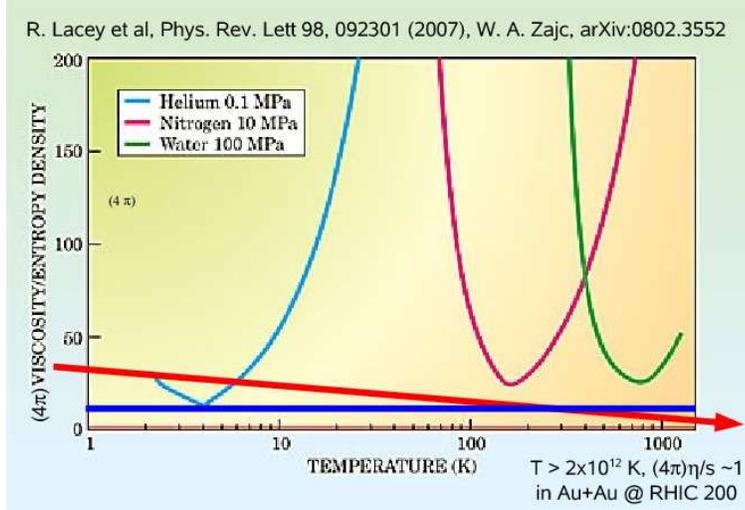} 
  \end{center}
\vspace*{-0.5cm}
  \caption{
Comparison of the properties of the perfect fluid created in 
$\sqrt{s_{NN} = 200 } $ GeV Au+Au collisions with less extraordinary materials
like water, nitrogen or helium. 
}
\label{f:HTCs}
\end{figure}

The following milestones lead to this important discovery:
PHENIX was the first to observe a {\it new phenomena} in 0-10 \% central Au+Au collisions
at $\sqrt{s_{NN}} = 130 $ GeV at RHIC:
the suppression of particle production with high transverse momentum, 
the first RHIC discovery
that made it to the cover page of The Physical Review Letters in January 2002.
However, it was not clear initially if this effect is due to the nuclear modification
of the structure functions (initial conditions) at such a high energies, or if this
is indeed a hadronic final state effect.
As a  control, $d+Au$ measurement was performed and all the four RHIC collaborations:
BRAHMS, PHENIX, PHOBOS and STAR reported the absence of suppression in these reactions.
This discovery implied that the suppression in Au+Au reaction is a final state effect,
due to the formation of {\it a new form of matter}, 
that also made its way to the cover page of the Physical Review Letters in August 2003.
The third milestone was the publication of the so called ``White Papers" or review papers
by all the four RHIC experiments. After several year's worth of high energy collisions,
and from a detailed analysis of the elliptic flow data, a consensus interpretation emerged
that the fireball made in Au+Au collisions at RHIC behaves like a liquid of strongly interacting
constituents, also known as ``the perfect fluid". This discovery became also known as
the Top Physics Story for 2005 by the American Institute of Physics.
This discovery has been considerably sharpened when STAR and  PHENIX pointed out that 
the observed elliptic flow patterns scale with the number of constituent quarks and
strange and even charm quarks participate in the flow. Although the theoretical interpretation
of this effect is still open for discussions in particular because the unsolved problem
of quark confinement in QCD prevents the application of first principle QCD calculations 
for this phenomena, in my opinion the experimental evidence is very clear, it is irrefutable that quark degrees of freedom are active and the perfect fluid seen in Au+Au collisions
 is a fluid of quarks~\cite{Adare:2006nq}.
(The role of gluons is less clear and less directly measurable 
from the experimental point of view.)
The fifth milestone was the quantification, how perfect is the perfect fluid at RHIC?
Answers were obtained by measuring the so called kinematic viscosity $\eta/s$,
which is the ratio of the shear viscosity to the entropy density.
Two theoretical analyses were published in 2007 based on elliptic flow patters,
a third measurement was based on the transverse momentum correlations, while PHENIX
studied the energy loss and flow of heavy (charmed) quarks and based
on a charm diffusion  picture, found that $\eta /s = (1.3 - 2.0) \frac{1}{4\pi}$
~\cite{Adare:2006nq}. Even more recently, PHENIX was able to put a lower limit
on the initial temperature of the fireball at RHIC from the analysis of direct photon
data~\cite{:2008fqa}, $T_i > 220 $MeV. These numbers can be compared
to similar characteristics of other known fluids, like water, liquid nitrogen or helium,
see Fig.~\ref{f:HTCs}, based on refs.~\cite{Lacey:2006bc,Zajc:2007ey}.

Note that $^4$He becomes superfluid at
extremely low temperatures and its kinematic viscosity $\eta/s$  
reaches a minimum at the onset of superfluidity, so for superfluid $^4$He
$\eta/s \ge 10 \frac{1}{4 \pi}$. The matter created 
in Au+Au collisions at RHIC  has temperatures larger than 2 Terakelvin,
nevertheless its kinematic viscosity is the lowest value ever produced
in laboratory: it is at least a factor of 4 smaller than that of superfluid $^4$He. 
We may thus refer this property of the matter created in Au+Au collisions at  RHIC  
as {\it high temperature superfluidity}~\cite{Csorgo:2008pe}: the matter created
in Au+Au collisions at RHIC is the most perfect fluid ever made by humans. 

We gain information on the type of transition from hadronic matter to quark matter
with the help of the Bose-Einstein correlations. 
By now, circumstantial evidence  is obtained
that this transition is either a cross-over or, 
a non-equilibrium transition.
This consensus opinion is based on important and highly selective  constraints given  
by Bose-Einstein correlations and particle interferometry data in Au+Au collisions 
at RHIC~\cite{Bekele:2007ee}.

\section{Conclusions}
After having discussed $e^+$ $e^-$, hadron-proton and heavy ion  reactions one after the other, 
and based on the presented evidence let me attempt to  give
answers to the questions discussed in the Introduction, keeping in mind that these
answers were worked out predominantly from the point of view of Bose-Einstein correlations
and their models in these reactions. 

{\it Can thermal \& hydrodynamical models describe 
$e^{+}$+$e^{-}$, $h$+$p$ and $A$+$B$ reactions?}
It seems that thermal models cannot interpret particle multiplicities, spectra and Bose-Einstein
correlations in $e^+$ + $e^-$ reactions at the present level of experimental precision, 
and using the conventional threshold for acceptable confidence levels (99.9 \% $\ge$ CL $\ge$ 0.1 \% ).
However, hydrodynamical and thermal models are remarkably successful in describing soft ($p_t < 1.5 $ GeV) hadron-proton, proton-proton and heavy ion reactions.
At higher values of the transverse momenta, jet physics and interaction of jets and the 
hydrodynamical medium opens up new research directions at the intersection of
particle and nuclear physics.

{\it What heavy ion physics can learn from 
$e^{+}$+$e^{-}$, $h^-$+$p$ and $p$+$p$ collisions?}
One lesson that I presented was to take statistical analysis and confidence level determinations seriously. Based on detailed and precision analysis of Bose-Einstein correlations in two-jet events
and a simultaneous analysis of single particle spectra, the time evolution, 
a movie or a video like film of particle emission has been 
already reconstructed in $e^+$ $e^-$ reactions at LEP. In heavy ion physics, only snapshot like pictures can be reconstructed at present. Further developments of the femtoscopic tools are needed
to allow for a video like reconstruction of the time evolution of particle emission in heavy ion reactions. Based on Bose-Einstein data in $e^+$ + $e^-$ reactions, it seems that the most probable
value of the proper-time parameter of particle production  is $\tau = 0.3$ fm/c, a surprisingly short value. It would be interesting to consider the phenomenological consequences of this number
in heavy ion reactions, and if possible, to extract similar numbers for jets that are produced
in a nuclear medium.

{\it How can correlations be used to determine the size of 
the interaction region and the characteristics of phase transitions?}
Of course a complete answer to this question goes well beyond the scope of this
conference contribution. Let me just emphasize here, that correlations are routinely  used
to take a snapshot picture of the interaction 
region~\cite{Csorgo:1999sj,Lednicky:2002fq,Lisa:2005dd,Bekele:2007ee}.
The resolution of these snapshot pictures has been increased recently and more detailed
information about structures (like a ring of fire) or heavy tails (non-Gaussian behavior) 
are seen in all kind of reactions~\cite{Bekele:2007ee}.
Recent progress even allowed for the determination  of the time evolution of
the region of particle production in $e^+$ $e^-$ reactions, based on a non-thermal
description. Similar techniques are not
yet developed for soft hadron-proton, proton-proton and heavy ion collisions,
where the thermodynamical and hydrodynamical models can readily be applied.
However, in heavy ion reactions matter formation and also a transition to a perfect
fluid of quarks has been experimentally proven (although with open theoretical issues).
Bose-Einstein correlations have been proven to constrain models in  an extremely efficient
manner. At present, models with a strong first order QCD phase transition or with a 
second order phase transition point disagree with Bose-Einstein correlation data 
in heavy ion collisions at RHIC, however, models with a cross-over transition or
with non-equilibrium rehadronization scenario cannot be 
excluded at present~\cite{Bekele:2007ee,Csorgo:2007iv}.

{\it Acknowledgments:} It is my pleasure to thank the Organizers for a 
most professionally organized conference. This research was supported by 
the OTKA grants  NK73143, T49466 as well as by a Senior Scholarship Award 
of the  Hungarian-American Enterprise Scholarship Fund.

\begin{footnotesize}
\bibliographystyle{ismd08} 
{\raggedright

}
\end{footnotesize}

\end{document}